\documentstyle[aps,psfig,epsf]{revtex}

\begin{document}

\title{\hfill quant-ph/9801077 \\[1cm]
SQUEEZED STATES OF A PARTICLE IN MAGNETIC FIELD}

\author{M. Ozana$^{(a)}$ and A.L. Shelankov$^{(a,b)}$}

\address{(a) Department of Theoretical Physics Ume\aa{\ }University, S-901
  87, Ume\aa, Sweden  \\
(b) Physico-Technical Institute,
  St. Petersburg, 194021, Russia }

\date{December 18, 1997; revised February 7, 1998}

\maketitle

\begin{abstract}
For a charged particle in a homogeneous  magnetic  field, we construct
stationary squeezed states which are eigenfunctions of the Hamiltonian
and the  non-Hermitian operator  $\hat{X}_{\Phi} = \hat{X} \cos \Phi +
\hat{Y} \sin \Phi$, $\hat{X}$ and $\hat{Y}$ being the  coordinates  of
the Larmor circle center and $\Phi$ is a  {\em complex} parameter.  In
the  family  of the squeezed  states, the  quantum uncertainty  in the
Larmor circle position is minimal.  The wave functions of the squeezed
states in the coordinate representation are found and their properties
are discussed.  Besides,  for arbitrary gauge of  the vector potential we
derive the symmetry operators of translations and rotations.
\end{abstract}

\bigskip

The problem of a charged quantum particle moving on a plane subject to
a homogeneous magnetic field is met in various physical contexts, and
it has been extensively studied in the literature and presented in
text-books \cite{LanLif77}.  A specific feature of the problem is that
the energy spectrum given by discrete Landau levels is multiple
degenerate: the number of linearly independent eigenstates belonging
to the $N$-th Landau level is proportional to the area of the plane
accessible to the particle.  The degeneracy is related to the
translational invariance: As a classical Larmor circle can be put
anywhere in the plane, a suitably defined operator of translation
$\hat{T}_{\bbox{a}}= e^{{i\over{\hbar}}\bbox{a\cdot}\hat{\bbox{P}}}$,
$\hat{\bbox{P}}$ being the generator of translations, commutes with
the Hamiltonian and upon acting on an energy eigenfunction
$\psi_{N}(\bbox{r})$ produces another eigenstate, shifted in space:
$|\hat{T}_{\bbox{a}}\psi_{N}(\bbox{r})|^{2}= |\psi_{N}(\bbox{r}+
\bbox{a}) |^{2}$ \cite{Zak,LifPit78}.  Non-collinear translations do
not commute in a magnetic field, and $[\hat{P}_{x}, \hat{P}_{y}]\neq
0$.  The existence of two non-commuting Hermitian operators,
$\hat{P}_{x}$ and $\hat{P}_{y}$, each of them commuting with the
Hamiltonian leads \cite{LanLif77} to the degeneracy.

The stationary wave functions corresponding to a degenerate Landau
level may be chosen to be eigenfunctions of either $\hat{P}_{x}$ or
$\hat{P}_{y}$ (but not both simultaneously).  The eigenvalues of
$\hat{P}_{x,y}$ are real, and the translation operator $e^{ia
\hat{P}_{x}}$ (or $e^{ia \hat{P}_{y}}$) applied to the corresponding
eigenfunction gives only an overall phase factor.  The modulus remains
unchanged by the translations, so that the eigenstates must be
infinitely extended in the $x-$ (or $y-$) direction.  The wave
functions of an electron in a magnetic field first found by Landau
\cite{LanLif77,Lan30} give an example: factorized as $e^{i px}\chi
(y)$, they are eigenfunctions of $\hat{P}_{x}$ and are infinitely
extended in the $x-$direction (strip-like states).

As discussed later, in the relation
\begin{equation}
\hat{\bbox{P}} = {e\over c} \bbox{B\times \hat{R}}\,, 
\label{kfa}
\end{equation}
$\hat{\bbox{R}}= (\hat{X}, \hat{Y})$ has the meaning of the operator 
corresponding to the classical coordinate of  the Larmor circle center
(the guiding center);
$\hat{X}=  {c\over e B} \hat{P}_{y}$ and $\hat{Y}=- {c\over e B}
\hat{P}_{x}$.
The variable $\bbox{R}$ has a simple classical interpretation, and for
this reason, it will be used below rather than $\bbox{P}$.

Instead of $\hat{X}$ or $\hat{Y}$ ($\hat{P}_{x}$ or $\hat{P}_{y}$),
one may choose their Hermitian linear combination $
\hat{X}_{\Phi}=\hat{X}\cos\Phi + \hat{Y}\sin \Phi $ with a real
$\Phi$.  The corresponding eigenstates are ``strips'' the orientation
of which depends on the angle $\Phi$.
A different class of states can be obtained if the wave
function is chosen to be an eigenfunction of the {\it non-Hermitian}
operator $\hat{X}_{\Phi }$ with a {\it complex} ``angle'' $\Phi =
\Phi_{1}+ i\Phi_{2}$. 
By virtue of the relation in Eq.(\ref{kfa}), the eigenfunctions of
$\hat{X}_{\Phi}$ are also eigenfunctions of $\hat{P}_{\Phi+
{\pi\over{2}}}$.  In the case of a general complex $\Phi $,
eigenvalues of the non-Hermitian operator $\hat{P}_{\Phi +
{\pi\over{2}}}$ are complex numbers, and the above argument concerning
an infinite extension of the state is not applicable; the
eigenfunctions turn out to be localized ({\it i.e.} the wave function
vanishes at infinity).
 
In the terminology of quantum optics (for a review see
\cite{ZhaFenGil90} and references therein) these states belong to the
class of {\em squeezed} states, generalization of the
coherent states.
In optics the squeezed state is defined as an
eigenfunction of a non-Hermitian operator, $\hat{x}- i \lambda
\hat{p}$, built of two non-commuting variables, the coordinate and
momentum of a harmonic oscillator ($\lambda$ being the squeeze
parameter). 
A distinctive feature of squeezed states is that the
quantum uncertainty in the non-commuting variables, is as minimal as
allowed by the uncertainty relation (minimum uncertainty states). 
The purpose of the paper to analyze
properties of the squeezed states, eigenfunctions of $\hat{X}_{\Phi}$.

Solutions to the Schr\"odinger equation for a charge in a magnetic
field corresponding to non-spreading wave packets with a classical
dynamics -- the coherent states in the modern terminology -- were
first built by Darwin as early as in 1928 \cite{Dar28}.  More
recently, the coherent states in the magnetic field problem have been
extensively studied by Malkin and Man'ko \cite{MalMan69}, and Feldman
and Kahn \cite{FelKah70} (see also
Ref.\cite{PavlovProkhorov,Rashba,Varro}). 
In the coherent states, the quantum
uncertainties in the $X$ and $Y$ coordinates of the Larmor center are
equal. 
 Various generalizations to the
squeezed states have been done by Dodonov {\it et al.}
 \cite{DodonovKurmyshevManko88,DodonovMankoPol94} 
and Aragone \cite{Aragone93}: 
In the general squeezed state, the uncertainty in one of the
coordinates is reduced at the expense of the other one so that their
product remains intact.

In the present paper, we consider {\em stationary} states,  building the
squeezed states from the energy eigenfunction belonging to a given
Landau level. 
Being stationary, these states are of different kind than the moving
squeezed wave packets of Ref. \cite{DodonovKurmyshevManko88,Aragone93}.

The paper is organized as follows. In Section \ref{symmetry}, we
review some general features of the quantum problem of a charge in a
magnetic filed.  In Section \ref{stuff}, we define the squeezed states
and explicitly find the wave functions in the coordinate
representation.  In Section \ref{prop}, properties of the squeezed
states are analyzed.  In the Appendix, we suggest a method which
allows one to construct symmetry operators for an arbitrary gauge of
the vector potential and apply the method to the case of a homogeneous
magnetic field.

\section{General Properties}\label{symmetry}

The Hamiltonian of a particle with mass $m$ and electric charge $e$
moving in the $x-y$ plane in a magnetic field reads
\begin{equation}
\hat{H}= {1\over{2m}}\left(\hat{\bbox{p}} - {e\over c} \bbox{A}\right)^{2},
\label{9aa}
\end{equation}
here the vector potential $\bbox{A}(A_{x}, A_{y})$
corresponds to a homogeneous magnetic field perpendicular to the plane, 
 $(\text{\bf rot }\bbox{A})_{z}= B$. 
The choice of signs in some of the below formulae depends on the sign
of $eB$; 
for definiteness, we assume  $eB>0$. 
In terms of the ladder
operators 
\begin{equation}
\hat{c}= {1 \over \sqrt{2}\,l\, \omega_c} \left( \hat{v}_x + i\hat{v}_y
\right) \;\; , \;\;  
\hat{c}^{\dagger}= {1 \over \sqrt{2}\, l \,\omega_c} \left( \hat{v}_x - i\hat{v}_y \right) 
\label{oba77}
\end{equation}  
where $\hat{v}_{x,y}$ are the non-commuting components of the velocity
operator  
\begin{equation}
 m\hat{\bbox{v}} = \hat{\bbox{p}}-{e\over c}\bbox{A} \;\; , \;\;
 [\hat{v}_{x}, \hat{v}_{y}]=i {\hbar^{2}\over{m^{2}l^{2}}}\; , 
\label{8aa}
\end{equation}
(the cyclotron frequency $\omega_{c}= {|eB|\over{mc}}$ and the magnetic
length 
$l= \sqrt{\hbar c/|eB|}$), the Hamiltonian Eq.(\ref{9aa}) can be
conveniently written as
\begin{equation}
\hat{H} = \hbar \omega_c \left( \hat{c}^{\dagger} \hat{c} +{1 \over
2}\right) \hspace{2ex}, \hspace{2ex} \left[ \hat{c} ,\hat{c}^{\dagger}
\right] = 1 \, . 
\label{7aa}
\end{equation}

In the presence of a homogeneous magnetic field, the translations in
the x-y plane and rotations around the $z-$axis remain symmetry
elements.  The reflection, $\sigma_{v}$, in a plane passing through
the $z-$axis ( $yOz$, for definiteness), reverses the magnetic field
and is not a symmetry transformation.  However, the product
$\sigma_{T}\equiv T \cdot \sigma_{v}$ of time-reversal $T$ and
$\sigma_{v}$ , both reversing the field, is a valid symmetry.

It is well-known that the Hamiltonian Eq.(\ref{9aa}) may not commute
with the operators associated with the physical symmetries because the
vector field $\bbox{A}(\bbox{r})$ has a lower symmetry than the
corresponding magnetic field.  If this is the case, the coordinate
transformation should be accompanied by a certain gauge transformation
which compensates the change in $\bbox{A}(\bbox{r})$ \cite{LifPit78}.
A procedure of constructing the transformation, which is valid for
arbitrary gauge of the vector potential, is presented in the Appendix.

As shown in the Appendix, the operators of finite translations,
$\bbox{r}\rightarrow \bbox{r}+ \bbox{a}$, are built of the generator
of translations
\begin{equation}
\hat{\bbox{P}}= \hat{\bbox{p}}- {e\over c} \bbox{A} + {e\over c} \bbox{B\times r}.
\label{jba}
\end{equation} 
The operator $\hat{\bbox{P}}$ commutes with the Hamiltonian
Eq.(\ref{9aa}) as manifestation of the translation invariance
preserved in a homogeneous magnetic field.  The components of
$\hat{\bbox{P}}$ obey the commutation relation
\begin{equation}
[\hat{P}_{x}, \hat{P}_{y}]= -i {e\over c} \hbar B\,\,.
\label{mba}
\end{equation}
Equation (\ref{jba}) is valid in an arbitrary gauge of the vector
potential $\bbox{A}$.  In case of the symmetric gauge, $\bbox{A}=
{1\over{2}} (\bbox{B\times r})$, Eq.(\ref{jba}) gives the expression
first found by Zak \cite{Zak}.

Presenting $\hat{\bbox{P}}$ in the form in Eq.(\ref{kfa}), one indeed
recognizes in $\hat{\bbox{R}}_{}$ the center of the Larmor circle (the
guiding center),
\begin{equation}
\hat{\bbox{R}}_{}= \hat{\bbox{r}} + {m c\over{e B^{2}}} \hat{\bbox{v}}
\bbox{\times B}, 
\label{lba}
\end{equation}
an integral of motion known from classical mechanics. 
The following commutation relations  
\begin{eqnarray}
  \left[\big(\hat{\bbox{R}}_{} \big)_{l}\;,\; 
\big(\hat{\bbox{P}}\big)_{m}\right]
  & =  &     i \hbar \delta_{lm}\label{qba1}\\   
 {[}\hat{X}, \hat{Y}{]} & =& 
{l^2\over i}
\label{qba2}
\end{eqnarray}
can be readily derived from Eqs.(\ref{mba}),  and (\ref{lba}).

\section{Squeezed states}\label{stuff}

From Eq.(\ref{qba2}), the $X$- and $Y$-coordinates of the Larmor
circle center are incompatible quantum variables.  Given the
commutator Eq.(\ref{qba2}), their variances obey the standard (see
e.g. \cite{Mes65}) uncertainty relation:
\[
(\Delta X)^{2}(\Delta Y)^{2} \ge {1\over{4}} l^{4}\; ,
\nonumber
\]
where the variance of a variable $A$ is defined as $(\Delta A)^{2}=
\left<\left(\Delta \hat{A}\right)^{2}\right>$ with $\Delta \hat{A}$
here and below standing for $\Delta \hat{A}= \hat{A}- \left< \hat{A}
\right>$.  If the uncertainty relation is satisfied with the equality
sign, it is said that the system is in a minimum uncertainty state or,
in other words, in a coherent or, more generally, in a squeezed state.

We construct the stationary {\em squeezed} state, $|\bbox{R}, N
\rangle $ as a simultaneous eigenfunction of the Hamiltonian
Eq.(\ref{9aa}) and the operator $\hat{X}_{\Phi}$,
\begin{equation}
\hat{X}_{\Phi}\equiv \hat{X} \cos \Phi  + \hat{Y}\sin \Phi \; ,
\label{vba}
\end{equation}
where
$\Phi $ is a complex parameter $\Phi = \Phi_{1}+ i \Phi_{2}$.  Under a
$\phi $-rotation around the $z$-axis, $\Phi $ transforms to $\Phi
\rightarrow \Phi'= \Phi - \phi $, and $\hat{X}_{\Phi}$ can be also
represented as
\begin{equation}
\hat{X}_{\Phi}= \hat{X}' \cosh \Phi_{2} + i \hat{Y}' \sinh \Phi_{2}
\label{wba}
\end{equation}
where $ \hat{X}'= \hat{X}\cos \Phi_{1} + \hat{Y}\sin \Phi_{1}$ and $
\hat{Y}'= -\hat{X}\sin \Phi_{1} + \hat{Y}\cos \Phi_{1}$ are the
Cartesian components of the guiding center $\bbox{R}$ in the principal
axes where $\Phi_{1}'=0$.

The state $ |\bbox{R}, N \rangle$ is found from the following system
of equations
\begin{eqnarray}
\hat{H} |\bbox{R}, N  \rangle    & =  & \hbar \omega_{c} (N +
     {1\over{2}})|\bbox{R}, N  \rangle \,\, ,   \label{oba1}\\   
\hat{X}_{\Phi}  |\bbox{R},N\rangle    & = &
(X\cos\Phi + Y\sin \Phi)|\bbox{R},N\rangle \,\,. 
\label{oba}
\end{eqnarray} 
The quantum numbers of a state are the Landau level number $N$, and
the expectation value of the guiding center position, $\bbox{R}(X,Y)$;
two real parameters $X$ and $Y$ specify the complex eigenvalue
$X_{\Phi}$.

The operator $\hat{X}_{\Phi}$ is not Hermitian and the eigenvalue
problem may or may not have solutions among physically admissible
normalizable functions, $\langle \bbox{R},N | \bbox{R},N \rangle=1
$. To find necessary conditions for the existence of physical
solutions, we note that Eq.(\ref{oba}) leads to $\langle \bbox{R},N
|\Delta \hat{X}_{\Phi}^{\dagger}\Delta \hat{X}_{\Phi} |
\bbox{R},N\rangle= 0$, or, using the representation in Eq.(\ref{wba})
and Eq.(\ref{qba2}),
\begin{equation}
 (\Delta X')^{2}+ \tanh^{2}\Phi_{2} (\Delta Y')^{2}= - l^{2}
 \tanh\Phi_{2}\,\,. 
\label{ofa}
\end{equation}
Observing that the l.h.s. is positive definite, we conclude that
Eq.(\ref{oba}) has normalizable solutions only if $\Phi_{2}<0$ (or,
more generally, $eB \Phi_{2} <0$ ).

The real and imaginary parts of the relation $\langle \bbox{R},N |
\left(\Delta \hat{X}_{\Phi} \right)^{2}|\bbox{R},N \rangle =0$, give
\begin{eqnarray}
(\Delta Y')^{2} \,\tanh^{2}\Phi_{2}&=&(\Delta X')^{2}
\label{nfa1}\\
 \left<\Delta \hat{X}' \Delta \hat{Y}'+\, \Delta \hat{Y}' \Delta
   \hat{X}'\right>  
&=& 0 \label{nfa2}\\
(\Delta X')^{2}(\Delta Y')^{2}& = &{l^{2}\over{4}}  \,\,. 
\label{nfa}
\end{eqnarray}
The last relation follows from the first two combined with
Eq.(\ref{ofa}). Also,
\begin{equation}
(\Delta X')^{2} = {l^{2}\over 2} |\tanh \Phi_{2}| \;\; , \;\;   
(\Delta Y')^{2} = {l^{2}\over 2} |\coth \Phi_{2}| \;\; . 
\label{sfa}
\end{equation}

According to Eq.(\ref{nfa}), the eigenfunctions of $\hat{X}_{\Phi}$
indeed belong to the class of minimum uncertainty states.  From
Eq.(\ref{nfa2}), one sees that the physical meaning of $\Phi_{1}$ is
to show the orientation of the principal axes, along which the quantum
fluctuations of the guiding center position are independent.  It
follows from Eq.(\ref{nfa1}) that $\Phi_{2}$ controls the relative
uncertainty of the projection of the guiding center onto the principal
axes.

To find the wave functions of the squeezed states,
 we first consider the states from the  Landau
level $N=0$, $|\bbox{R},0\rangle $ and solve the
following system of equations
\begin{eqnarray}
  \hat{c}\,|\bbox{R},0\rangle    & =  & 0        \;\; ,      \label{uba1}\\   
\hat{X}_{\Phi}|\bbox{R},0\rangle & =    & X_{\Phi } |\bbox{R},0\rangle \,\,.      
\label{uba}
\end{eqnarray}
(Eq.(\ref{uba1}) is equivalent to Eq.(\ref{oba1}) for the ground state $N=0$.)
To find the explicit form of the wave functions, we choose 
 the symmetric gauge
\begin{equation}
\bbox{A} = {1\over{2}}\ \bbox{B\times r} \, ,
\label{sba}
\end{equation}
where different directions are treated on equal footing. 

In the notations 
\begin{eqnarray}
\tilde{x}_{\Phi}    & =   & (x-X) \cos \Phi + (y-Y) \sin \Phi   \ ,
\nonumber\\    
\tilde{y}_{\Phi}   & =  &  - (x-X) \sin \Phi + (y-Y)
\cos \Phi \ , 
\label{tba}
\end{eqnarray}
($\tilde{y}_{\Phi} =\tilde{x}_{\Phi + {\pi\over{2}}} $), the operators
in Eq.(\ref{uba1}) and Eq.(\ref{uba}) take the form
\begin{eqnarray}
i {\sqrt{2}\over l}e^{-i\Phi} \hat{c} 
&=&  {\partial\over{\partial \tilde{x}_{\Phi} }}
  + i {\partial \over \partial \tilde{y}_{\Phi}}  + {1 \over 2 l^{2}} 
 \big( \tilde{x}_{\Phi}+i\tilde{y}_{\Phi} \big) + {1 \over 2 l^{2}} 
 \big( X_{\Phi}+i Y_{\Phi} \big) 
\\
\hat{X}_{\Phi} &=& \tilde{x}_{\Phi} - i l^2 {\partial \over \partial
  \tilde{y}_{\Phi}} + X_{\Phi} \,\,, 
\label{M77ia}
\end{eqnarray}
where $Y_{\Phi}  =   - X \sin \Phi + Y \cos \Phi $.

In the coordinate representation, Eqs.(\ref{uba1},\ref{uba}) become a
system of linear differential equations of the first order for $\Psi
(\bbox{r}|0, \bbox{R})= \langle \bbox{r} |\bbox{R},0\rangle$.  The
normalized solution reads
\begin{equation}
\Psi(\bbox{r} \mid 0, \bbox{R} ) =  C_{\Phi} \exp{\left( -{1 \over 
2l^2} \left( \tilde{x}_{\Phi}^2 
+ i \tilde{x}_{\Phi} \tilde{y}_{\Phi} + i
\tilde{x}_{\Phi} Y_{\Phi} 
- i\tilde{y}_{\Phi } X_{\Phi} 
\right) \right)}, 
\label{Mzda}
\end{equation}
where $C_{\Phi }$ is the normalization constant
\begin{equation}
|C_{\Phi}|^{2}= {1 \over {2 \pi l^2}}
\sqrt{1 - \exp \, {2i(\Phi^* - \Phi)}} \ .
\label{Nwba}
\end{equation}
As expected, the function in Eq.(\ref{Mzda}) is normalizable and
Eq.(\ref{Nwba}) is meaningful, only if ${\em Im} \,\Phi <0$.

The normalized states for the N-th Landau level, eigenfunctions of the
Hamiltonian in Eq.(\ref{7aa}), can be now found from
\begin{equation}
\Psi(\bbox{r} \mid N, \bbox{R} )= {1\over{\sqrt N!}}
\big(\hat{c}^{\dagger} \big)^{N}\Psi(\bbox{r} \mid 0, \bbox{R} ) \,\, .
\label{yba}
\end{equation}

The calculation can be easily done with the help of the following
identity
\begin{eqnarray}
\hat{c}^{\dagger}& = &e^{-i\Phi } e^{\Lambda} \left({l \hbar
     \over{i\sqrt2}} {\partial\over{\partial \tilde{x}_{\Phi }}}
     \right) e^{- \Lambda } \,\, ,\nonumber\\   
  \Lambda      &  \equiv  &
{1\over{2l^{2}}}\big(\tilde{x}_{\Phi}^{2}
- i \tilde{x}_{\Phi} \tilde{y}_{\Phi } - i \tilde{x}_{\Phi} Y_{\Phi} + i 
\tilde{y}_{\Phi} X_{\Phi}\big) \,\, .
\label{1ba}
\end{eqnarray}
After some algebra we obtain, 
\begin{equation}
\Psi (\bbox{r}|N, \bbox{R}) = C_{\Phi
  }{\left(i e^{-i\Phi } \right)^{N}\over{\sqrt{2^{N}N!}}} 
\exp\, \left( {i{e \over 2 \hbar c}\bbox{B} \cdot (\bbox{R} \times \bbox{r})}
 -{1\over{2l^{2}}}
\tilde{x}_{\Phi} \left( \tilde{x}_{\Phi}
+ i\tilde{y}_{\Phi} \right)
\right)  H_{N}\left({\tilde{x}_{\Phi}\over{l}}\right) 
\label{3ba}
\end{equation}
where $H_{N}(\xi)$ is the Hermite polynomial, $H_{N}(\xi)=(-1)^{N}
e^{\xi^{2}} {d^{N} e^{-\xi^{2}}\over{d\xi^{N}}}$.  In the coordinate
representation, this expression gives the wave function of the
squeezed state $|\bbox{R},N \rangle $ centered at $\bbox{R}$ and
belonging to the $N$-th Landau level (in the symmetric gauge
Eq.(\ref{sba})); the ``rotation angle'' $\Phi $ in Eq.(\ref{tba}) is a
complex parameter, ${\em Im}\, \Phi < 0 $.

\section{Properties}\label{prop}

The basic features of squeezed states can be seen in Figs. \ref{Fig1}
and \ref{Fig2}, where the density and the current are plotted for a
typical state: $N=2$, $\Phi = i \Phi_{2}$, $|\tanh \Phi_2| =0.1$.  The
squeezed state is {\em localized} in the sense that it has a finite
extension in both $x-$ and $y-$directions.  Qualitatively, the
squeezed state is a superposition of classical Larmor orbits of radius
$\rho_{N} = \sqrt{2E_N \over m \omega^2}$, the centers of which are
positioned in the vicinity of $\bbox{R}$ with a typical deviation
$\Delta X'$ and $\Delta Y'$. From Eq.(\ref{sfa}), $\Delta X'/\Delta Y'
= |\tanh \Phi_{2}| < 1 $ so that the state is elongated in the
direction of the principal $y'$-axis.  When $\Phi_{2}$ tends to zero,
the elongation increases and the squeezed state asymptotically
transforms into a ``strip'' (of length $\sim l/ |\Phi_{2}|$).

The wave function of a squeezed state from the N-th Landau level has
$N$ {\it isolated zeroes} in the $x-y$ plane. The zeroes are at the
points on the line $Im\;\tilde{x}_{\Phi }=0$, where the Hermite
polynomial $H_{N}({\tilde{x}_{\Phi }\over l})$ has its $N$ roots.  In
the limit $\Phi_{2}\rightarrow -\infty$, the zeroes gather together at
the point $\bbox{r}= \bbox{R}$.  This limit, where $\Delta X= \Delta Y
= {l\over 2 }$, gives the stationary coherent state introduced by
Malkin and Man'ko\cite{MalMan69}, the angular momentum eigenstate with
the eigenvalue $L_{z} = - \hbar N$.  In the coherent states, the
probability density is rotationally invariant.

In quantum optics, squeezed states can be presented as the result of
the action of the ``squeezing operator'' on the coherent state.  
Similar to \cite{DodonovKurmyshevManko88},
squeezing of the cylindrically symmetric coherent states
($Im \, \Phi \rightarrow -\infty $) is achieved by applying 
\begin{equation}
\hat{S} = \exp\left\{ {i \over 2l^2} r \Big( \hat{X}' \hat{Y}' + \hat{Y}'
\hat{X}' \Big) \right\} \, ,
\label{M6ia}
\end{equation} 
where $\tanh r = e^{2 \Phi_{2}}$.
In optics there are certain non-linear processes 
with the evolution operator in the form of $\hat{S}$ 
\cite{WallsMilburn94}. 
A coherent state then evolves into the squeezed state.
For the case of a particle in a magnetic field, 
the analogous problem of preparation of a squeezed state 
has been considered in \cite{DodonovMankoPol94}.

As discussed in the Appendix,
the product of the mirror and time reversal transformations is a valid
symmetry element.  As a consequence, the
distribution of the density and current are mirror symmetric (relative
to the principal axes) as also apparent in
Figs.(\ref{Fig1}-\ref{Fig4}).

Within a given Landau level, the squeezed states, eigenfunctions of a
non-Hermitian operator $\hat{X}_{\Phi}$, are {\it non-orthogonal}.
For the states defined in Eq.(\ref{3ba}) with the real positive
normalization constant $C_{\Phi }$ Eq.(\ref{Nwba}), the overlap
integral reads
\begin{eqnarray}
\langle \bbox{R};\Phi, N \mid \bbox{R}'; \Phi', N' \rangle &=&
\delta_{NN'} 
{
\Big(1 - \exp \big( {2i ( \Phi^{*} -\Phi )}\big)\Big)^{1/4}
\Big(1 - \exp
\big({2i ( \Phi'^{*} -\Phi')}\big) \Big)^{1/4}  
  \over \Big(1 - \exp \big({2i ( \Phi^{*} -\Phi' )}\big)\Big)^{1/2} 
}
\times \nonumber \\
 & \times & \exp \left( {{ i e \over 2\hbar c}\bbox{B}\cdot(  
\bbox{R}' \times \bbox{R})} +{i \over 2l^2}
{\left( X_{\Phi'}' - X_{\Phi'} \right) \left(X_{\Phi^*}' - 
X_{\Phi^*} \right)  \over \sin \left(\Phi' - \Phi^*\right)}   \right)
\label{M3da}
\end{eqnarray}
($Re\, \Big(1 - \exp \big({2i ( \Phi^{*} -\Phi')}\big)\Big)^{1/2} > 0$). 
The overlap of the states differing in the Larmor center position
$\bbox{R}$ or the parameter $\Phi $ does not depend on the Landau
level number $N$ as it follows from the Eq.(\ref{yba}) and the
commutation relation in Eq.(\ref{7aa}). Therefore, the overlap
integral in Eq.(\ref{M3da}) can be calculated using the Gaussian wave
functions for $N=0$ in Eq.(\ref{Mzda}).

Repeating the derivation known in the theory of coherent states of a
harmonic oscillator (see e.g. \cite{ZhaFenGil90}), one can show that
the set of squeezed states is complete {\it i.e.} the closure
relation,
\begin{equation}
\hat{1} = \sum \limits_{N=0}^{\infty} \int {d^{2}\bbox{R}\over 2\pi l^2}
\,  |\bbox{R}, N \rangle \langle \bbox{R}, N | \,\, , 
\label{M6iaa}
\end{equation}
is valid. As in the harmonic oscillator case, the states $|\bbox{R},N
\rangle $ with continuously varying $\bbox{R}$ form an overcomplete
set within the $N$-th Landau level.  Repeating Perelomov's arguments
\cite{a9}, one can show that the subset of states with $\bbox{R}$'s on
the sites of a periodic lattice is overcomplete if the lattice is too
dense (the unit cell area $s_{0}< 2\pi l^{2}$) and is not complete for
a too dilute lattice ( $s_{0}> 2 \pi l^{2}$).  When $s_{0}= 2 \pi
l^{2}$ ({\it i.e.}  the flux through the unit cell equals to the flux
quantum ${ h c\over e }$), the system of the functions is complete and
it remains complete even if a {\em single} state is removed; it
becomes incomplete, however, if any {\em two} states are removed.

In conclusion, stationary squeezed states of a charged
particle  in a homogeneous magnetic field have been constructed and analyzed.
The distinctive
feature of the squeezed states is the minimal quantum uncertainty of
the position of the Larmor circle center. The family of the squeezed
states is characterized by the squeezing parameter $Im \,\Phi$
variation of which allows one to transform gradually strip-like states
(infinitely extended in the direction controlled by $Re \,\Phi $) to
eigenfunctions of the angular momentum with rotationally invariant
density distribution.  The squeezed states have a rather simple
structure: As it follows from Eqs.(\ref{3ba} and \ref{tba}), a general
squeezed state can be obtained from a Landau ``strip'' by a complex
angle rotation of the coordinates. The simplicity of the construction
gives the hope that the squeezed states may turn out to be useful.

We are grateful to J{\o}rgen Rammer for reading the manuscript and
valuable remarks. 
We also thank V.V. Dodonov for drawing our
attention to the references
\cite{DodonovKurmyshevManko88,DodonovMankoPol94,Aragone93}. 
This work was supported by the Swedish Natural
Science Research Council.

\appendix 

\section{Symmetries operators}
\label{append1}

The physical symmetry of a system in an external magnetic field is
controlled by the symmetry of the vector field $\bbox{B}(\bbox{r})$
(among other factors).  However, the Hamiltonian, $\hat{H}=
\hat{H}_{\bbox{A}}$ contains the vector potential $\bbox{A}(\bbox{r})$
rather than $\bbox{B}$.  The vector field $\bbox{A}(\bbox{r})$ has a
lower symmetry and, moreover, the spatial symmetry of
$\bbox{A}(\bbox{r})$ is gauge dependent.  For this reason, the
Hamiltonian often does not commute with the operator corresponding to
a physical symmetry element. A homogeneous magnetic field gives a
simple example: the translation invariance is preserved but the vector
potential is always $\bbox{r}$-dependent.  It is well-known
\cite{LifPit78} that the symmetry operator should include a certain
gauge transformation which compensates the change of the
$\bbox{A}$-field generated by the symmetry transformation.

To build the modified operators on a regular basis, we suggest the
following procedure.  For any given symmetry element ${\cal O}$, it is
always possible to find field $\bbox{A}^{({\cal O})}(\bbox{r})$,
$\text{\bf rot }\bbox{A}^{({\cal O})}= \bbox{B}$, which is invariant
relative to ${\cal O}$: ${\cal O}\bbox{A}^{(\cal{O})}=
\bbox{A}^{(\cal{O})}$.  Whatever gauge is chosen for $\bbox{A}$ in
Eq.(\ref{9aa}), the gauge transformation,
$\hat{G}_{\cal{O}}^{-1}\hat{H}_{\bbox{A}}\hat{G}_{\cal{O}}$, specified
by
\begin{equation}
\hat{G}_{\cal{O}} = e^{i \chi^{(\cal{O})}}\;\; , \;\;  
\chi^{(\cal{O})}(\bbox{r})= {e\over{\hbar c }} \int\limits_{0}^{\bbox{r}} \, d\bbox{r}\,
\big(
\bbox{A} - \bbox{A}^{({\cal O})}
\big)  \;\; , 
\label{dba}
\end{equation}
changes the vector potential entering 
the Hamiltonian  from
$\bbox{A}$ to  $ \bbox{A} - \bbox{\nabla }_{\bbox{r}}\chi^{({\cal O})}=
\bbox{A}^{({\cal O})}$, {\it i.e.} 
$\hat{G}_{\cal{O}}^{-1}\hat{H}_{\bbox{A}}\hat{G}_{\cal{O}}=
\hat{H}_{\bbox{A}^{({\cal O})}}$ . By construction, $\cal{O}$ does
not change the vector field $\bbox{A}^{(\cal{O})}(\bbox{r})$ and,
therefore, 
the transformed operator $\hat{G}_{\cal{O}}^{-1}\hat{H}_{\bbox{A}}\hat{G}_{\cal{O}}$
commutes with $\hat{\cal{O}}$,
{\it i.e.} 
\[
\left(
\hat{G}_{\cal{O}}^{-1}\hat{H}_{\bbox{A}}\hat{G}_{\cal{O}}
\right)
\hat{\cal{O}}=
\hat{\cal{O}}
\left(
\hat{G}_{\cal{O}}^{-1}\hat{H}_{\bbox{A}}\hat{G}_{\cal{O}}\right)
\,\,\,\,\, 
\text{or}
\,\,\,\,\,
\hat{H}_{\bbox{A}}
\left(
\hat{G}_{\cal{O}} \hat{\cal{O}}\hat{G}_{\cal{O}}^{-1} 
\right)
=
\left(
\hat{G}_{\cal{O}} \hat{\cal{O}}\hat{G}_{\cal{O}}^{-1} 
\right)
\hat{H}_{\bbox{A}} \,\,\, .
\] 
Therefore, it is the
operator
\begin{equation} 
\hat{\cal{O}}_{\bbox{A}} =
\hat{G}_{\cal{O}} \hat{\cal{O}}\hat{G}_{\cal{O}}^{-1} \;\; , 
\label{eba}
\end{equation}
which commutes with the Hamiltonian $\hat{H}_{\bbox{A}}$ and
represents the symmetry element $\cal{O}$. As such, the operator
$\hat{\cal{O}}_{\bbox{A}}$ depends on the gauge chosen for the vector
potential $\bbox{A}$ in the Hamiltonian, but its matrix elements are
gauge invariant if the sandwiching functions are gauge transformed in
the usual manner.

Equivalently, the symmetry operator in Eq.(\ref{eba}) can be written
as
\begin{eqnarray}
\hat{\cal{O}}_{\bbox{A}}    & =  & 
e^{i \big(
\chi^{(\cal{O})}(\bbox{r})
-
\chi^{(\cal{O})}(\hat{\cal{O}}\bbox{r}) 
\big) }\  \hat{\cal{O}} \label{fba1}
\\   
     & =  &
\hat{\cal{O}}\  
e^{i \big( 
\chi^{(\cal{O})}(\hat{\cal{O}}^{-1}\bbox{r}) 
-
\chi^{(\cal{O})}(\bbox{r})
\big) } \, .
\label{fba}
\end{eqnarray}

Below, we analyze only the case of a homogeneous magnetic field but
the method is generally applicable.

In the presence of a homogeneous magnetic field, the translations in
the x-y plane and rotations around the $z-$axis remain (continuous)
symmetry elements.  As the reflection, $\sigma_{v}$, in a plane
passing through the $z-$axis ( $yOz$, for definiteness), reverses the
magnetic field, $\sigma_{v}$ is not a symmetry element. We note here
that the product of time-reversal $T$ and $\sigma_{v}$ , both
reversing the field, is a symmetry element. We denote the product by
$\sigma_{T}\equiv T \cdot \sigma_{v}$.

The transformation $\cal{O}$ changes the wave function as, $\psi
\stackrel{\cal{O}}{\longrightarrow} \hat{\cal{O}}\psi $, where
$\hat{\cal{O}}$ denotes the operator associated with $\cal{O}$.  For
translations and rotations,the form of the operator $\hat{\cal{O}}$ is
obvious; in the case of $\sigma_{T}$,
\begin{equation}
\hat{\sigma_{T}}\psi (x,y)=
\psi^{*}(-x, y) \, ,
\label{Mwia}
\end{equation}
and $\hat{\sigma_{T}}$ is an anti-linear anti-unitary operator.  From
$\hat{{\cal O}}^{-1}\hat{H}_{\bbox{A}}\hat{{\cal O}}\equiv
\hat{H}_{\cal{O}\bbox{A}}$, the transformed vector field
$\cal{O}\bbox{A}$ can be found. Again, the result is obvious in case
of rotations and translations. Under $\sigma_{T}$, which is the mirror
reflection of the polar vector field $\bbox{A}(\bbox{r})$ in
combination with time reversal ($\bbox{A}\rightarrow - \bbox{A}$), the
vector potential transforms as: $\hat{\sigma_{T}}A_{x}(x,y)=
A_{x}(-x,y)$, $\hat{\sigma_{T}}A_{y}(x,y)= -A_{y}(-x,y)$.

First, we consider the operator of a finite translation $\hat{\cal O}=
\hat{T}_{\bbox{a}}$, $\hat{T}_{\bbox{a}}\psi (x,y)= \psi (x+ a_{x},
y+a_{y})$, $\bbox{a}$ being the translation vector.  The vector
potential, $\bbox{A}^{(\cal{O})} \equiv \bbox{A}^{(\bbox{a})}$,
invariant relative to the translation along $\bbox{a}$,
$\bbox{A}^{(\bbox{a})}(\bbox{r})=\bbox{A}^{(\bbox{a})}(\bbox{r}+
\bbox{a}) $, can be taken as,
\begin{equation}
\bbox{A}^{(\bbox{a})}= \bbox{n} \Big((\bbox{B\times r}) \bbox{\cdot n}
\Big) \;\; , \;\;  \bbox{a}=
|\bbox{a}|\bbox{n}\;\; , 
\label{ega}
\end{equation}
(``Landau gauge''). 
Using Eq.(\ref{fba1}), the operator of magnetic translation reads
\begin{equation}
\hat{T}_{\bbox{A}}^{(\bbox{a})}=
e^{i{e\over{c\hbar}}\int\limits_{\bbox{r}+ \bbox{a}}^{\bbox{r}}d
  \bbox{r}'\Big(\bbox{A}(\bbox{r}')-
  \bbox{A}^{(\bbox{a})}(\bbox{r}')\Big) }
e^{{i\over{\hbar}}\hat{\bbox{p}}\bbox{\cdot a}},
\label{gba}
\end{equation}
$\hat{\bbox{p}}= {\hbar \over{i}}\bbox{\nabla }$ being the canonical
momentum.  As a consequence of the physical translational invariance,
this operator commutes with the Hamiltonian Eq.(\ref{9aa}) for
arbitrary gauge of the vector potential $\bbox{A}(\bbox{r})$.

Up to terms linear in $\bbox{a}$, $ \hat{T}_{\bbox{A}}^{(\bbox{a})} =
1 + {i \over \hbar} \bbox{a} \cdot \left( \hat{\bbox{p}} - {e \over c}
\bbox{A} + {e \over c} \bbox{A}^{(\bbox{a})} \right)$ or, using
Eq.(\ref{ega}), $\hat{T}_{\bbox{A}}^{(\bbox{a})} = 1 + {i \over \hbar}
\bbox{a} \cdot \left( \hat{\bbox{p}} - {e \over c} \bbox{A} + {e \over
c} \bbox{B\times r} \right) $. From here, one reads off the expression
for the generator of translations $\hat{\bbox{P}}$ and derives
Eq.(\ref{jba}).

This derivation of Eq.(\ref{jba}) links $\hat{\bbox{P}}$ to the
translational symmetry. The same expression for $\hat{\bbox{P}}$ can
be derived in a more intuitive manner: First, one considers the
operator of the guiding center $\hat{\bbox{R}}$, the expression for
which in Eq.(\ref{lba}) can be guessed from the correspondence
principle. Since $\bbox{R}$ is a classical integral of motion,
$\hat{\bbox{R}}$ must commute with the Hamiltonian.  Now, one {\em
defines} $\hat{\bbox{P}}$ by Eq.(\ref{kfa}) and comes immediately to
Eq.(\ref{jba}).

Next we consider rotations {\it i.e.}  ${\cal O}= R$.  The ``bare''
operator of a rotation around the $z$-axis is $\hat{R}=e^{{i\over
\hbar} (\bbox{r} \times \hat{\bbox{p}})_{z} \phi} $, $\phi$ being the
angle of the rotation.  The $R-$invariant vector potential is
$\bbox{A}^{(R)}= {1 \over 2} \bbox{B} \times \bbox{r}$. Applying
Eq.(\ref{eba}), the symmetry operator reads:
\begin{equation}
\hat{R}_{\bbox{A}}(\phi) = e^{i {e \over c\hbar}
\int\limits_{0}^{\bbox{r}} \, d\bbox{r}\,  
\big(
\bbox{A} - \bbox{A}^{(R)}
\big)}   
e^{{i\over \hbar} (\bbox{r} \times \hat{\bbox{p}})_{z}  \phi}  
 e^{-i {e \over c\hbar}
\int\limits_{0}^{\bbox{r}} \, d\bbox{r}\,  
\big( \bbox{A} - \bbox{A}^{(R)} \big)} \, . 
\label{M0ia}
\end{equation}
In the limit $\phi \rightarrow 0$,
$\hat{R}_{\bbox{A}}(\phi) \approx 1 + {i \over \hbar} \hat{{\cal L}}_{z}
\phi$, where  the generator of rotations $\hat{{\cal L}}_{z}$ is  
\[
\hat{{\cal L}}_{z}=
 e^{i {e \over c\hbar}
\int\limits_{0}^{\bbox{r}} \, d\bbox{r}\,  
\big(
\bbox{A} - \bbox{A}^{(R)}
\big)}   
(\bbox{r\times}\hat{\bbox{p}})_{z}
 e^{-i {e \over c\hbar}
\int\limits_{0}^{\bbox{r}} \, d\bbox{r}\,  
\big( \bbox{A} - \bbox{A}^{(R)} \big)} 
\,\, \, .
\]
Simplifying and using the expression for $\bbox{A}^{(R)}$, one gets
\begin{equation}
\hat{{\cal L}}_{z}= 
\left(\bbox{r} \times  \left(\hat{\bbox{p}} - {e\over c} \bbox{A}\right)\right)_{z} 
+ {e \over 2c}
\big(
\bbox{r} \times (\bbox{B} \times \bbox{r}) \big)_{z} \;\; .
\label{Mija1}
\end{equation}
For the case of a homogeneous magnetic field, this operator commutes
with the Hamiltonian for arbitrary gauge of the vector potential and
reduces to the usual angular momentum $\hat{L}_{z}=
\bbox{r\times}\hat{\bbox{p}}$ when $\bbox{A}, \bbox{B} \rightarrow 0$.
Also, $\hat{{\cal L}}_{z}= \hat{L}_{z}$ in the symmetric gauge
$\bbox{A}= {1\over 2 } \bbox{B\times r}$.

The operator ${\cal L}_{z}$ can be written in the following
identically equivalent forms
\begin{eqnarray}
\hat{{\cal L}}_{z}     & =  &    
{1 \over 2}{eB \over c}
\hat{\bbox{R}}^2 - {mc \over eB}\hat{H} \,
          \label{fga1}\\   
 \hat{{\cal L}}_{z}    &  = & 
{1 \over 2}{c \over eB}
\hat{\bbox{P}}^2 - {mc \over eB}\hat{H} ;
\label{fga}
\end{eqnarray}
$\hat{H}$, and the (two-dimensional) vectors $\hat{\bbox{P}}$ and
$\hat{\bbox{R}}$ are defined in Eqs.(\ref{9aa}), (\ref{jba}) and
(\ref{lba}), respectively.  One sees that the integral of motion
${\cal L}_{z}$ is, actually, a function of the other conserving
quantities $\bbox{P}$ (or $\bbox{R}$) and the energy $H$.

If the axis of the a rotation is shifted from the origin to the point
$\bbox{R}_{0}$, the generator of the rotations denoted as $\hat{{\cal
L}}_{\bbox{R}_{0} , z}$ reads
\begin{equation}
\hat{{\cal L}}_{\bbox{R}_{0} , z} ={1 \over 2} {eB
  \over c} \left( 
\hat{\bbox{R}} - \bbox{R}_{0} \right)^2 - {mc \over eB} \hat{H} \, .
\label{Maja}
\end{equation}

Note that the vector potential enters ${\cal L}_{z}$ Eq.(\ref{Mija1})
and the generator of translations $\hat{\bbox{P}}$ Eq.(\ref{jba}) only
in the gauge covariant combination $\hat{\bbox{p}} - {e\over c}
\bbox{A}$ so that their matrix elements are gauge invariant.

\begin{figure}[htbp] 
\centerline{
        \epsfxsize=16.0cm
        \epsfxsize=14.0cm
        \epsfbox{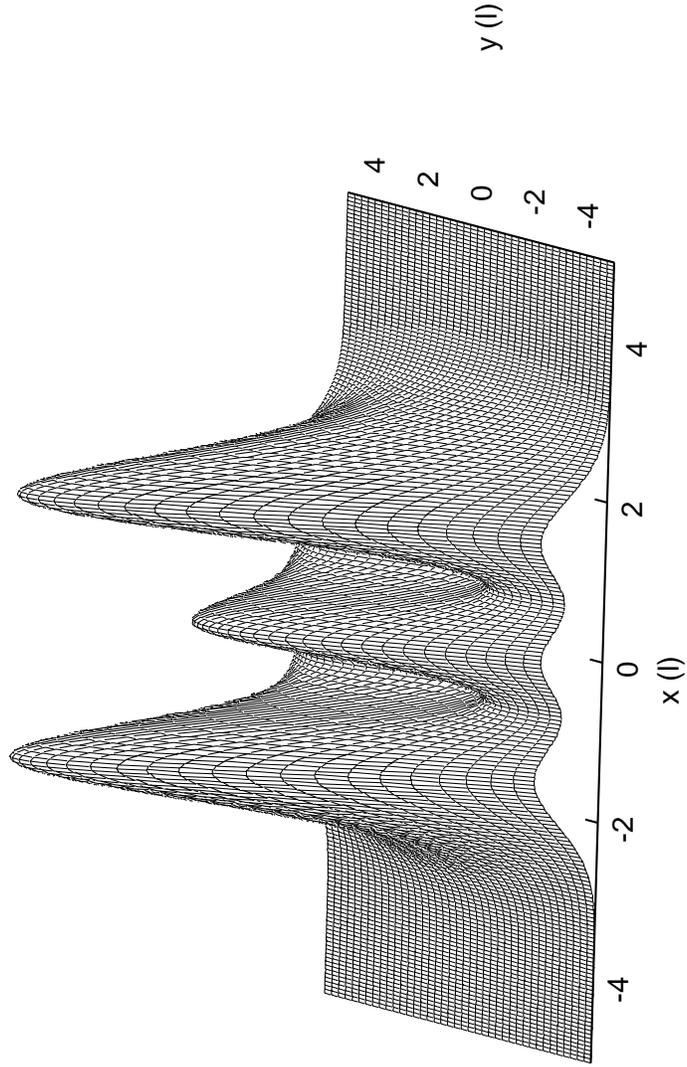}
        \vspace*{0.5cm}
        }
\caption{The probability density for the state $N=2$ ; $|\tanh \Phi_2|
  =0.1, \Phi_{1}= 0$ located at the origin $\bbox{R} = 0$. }
\label{Fig1}
\end{figure}

\begin{figure}
\centerline{
        \epsfxsize=16.0cm
        \epsfxsize=14.0cm
        \epsfbox{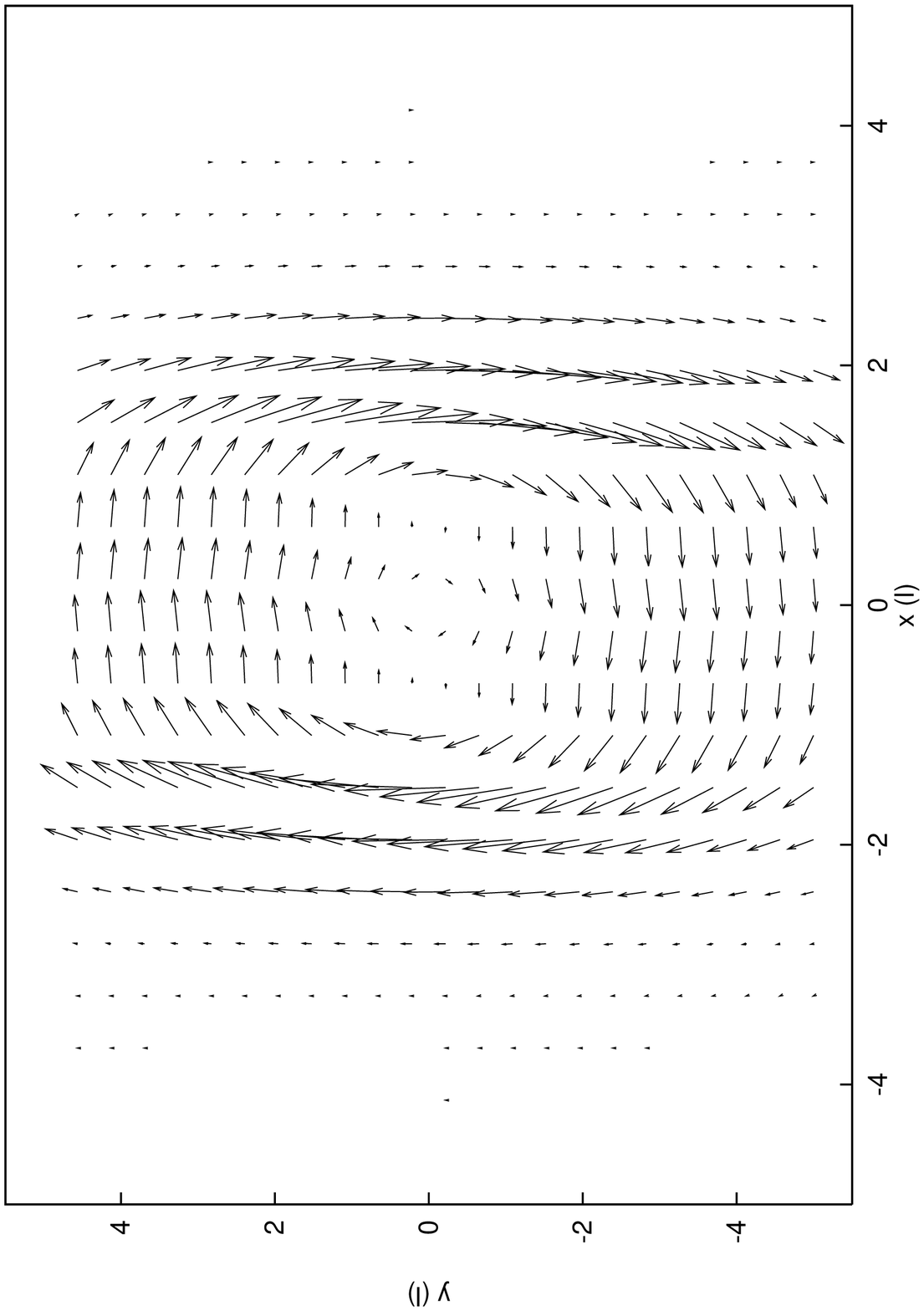}
        \vspace*{0.5cm}
        }
\caption{The current density for the state $N=2$ ; $|\tanh \Phi_2|
  =0.1, \Phi_{1}= 0$ located at the origin $\bbox{R} = 0$.} 
 \label{Fig2}
\end{figure}

\begin{figure}
\centerline{
        \epsfxsize=16.0cm
        \epsfxsize=14.0cm
        \epsfbox{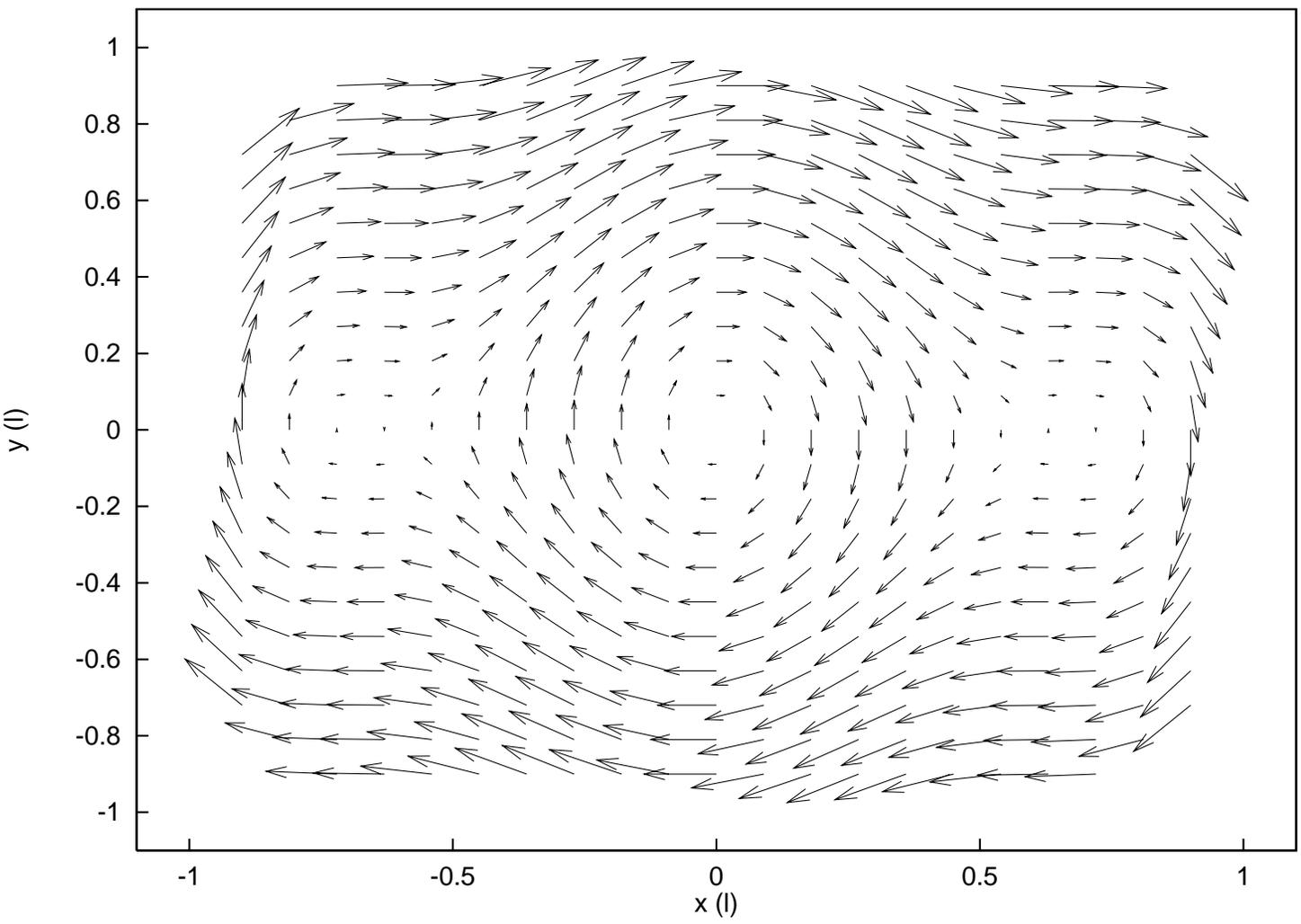}
        \vspace*{0.5cm}
        }
\caption{Detail of the current density for the state $N=2$ ; $|\tanh
  \Phi_2| =0.1, \Phi_{1}= 0$ located at the origin $\bbox{R} = 0$. }
 \label{Fig3}
\end{figure}

\begin{figure}
\centerline{
        \epsfxsize=16.0cm
        \epsfxsize=14.0cm
        \epsfbox{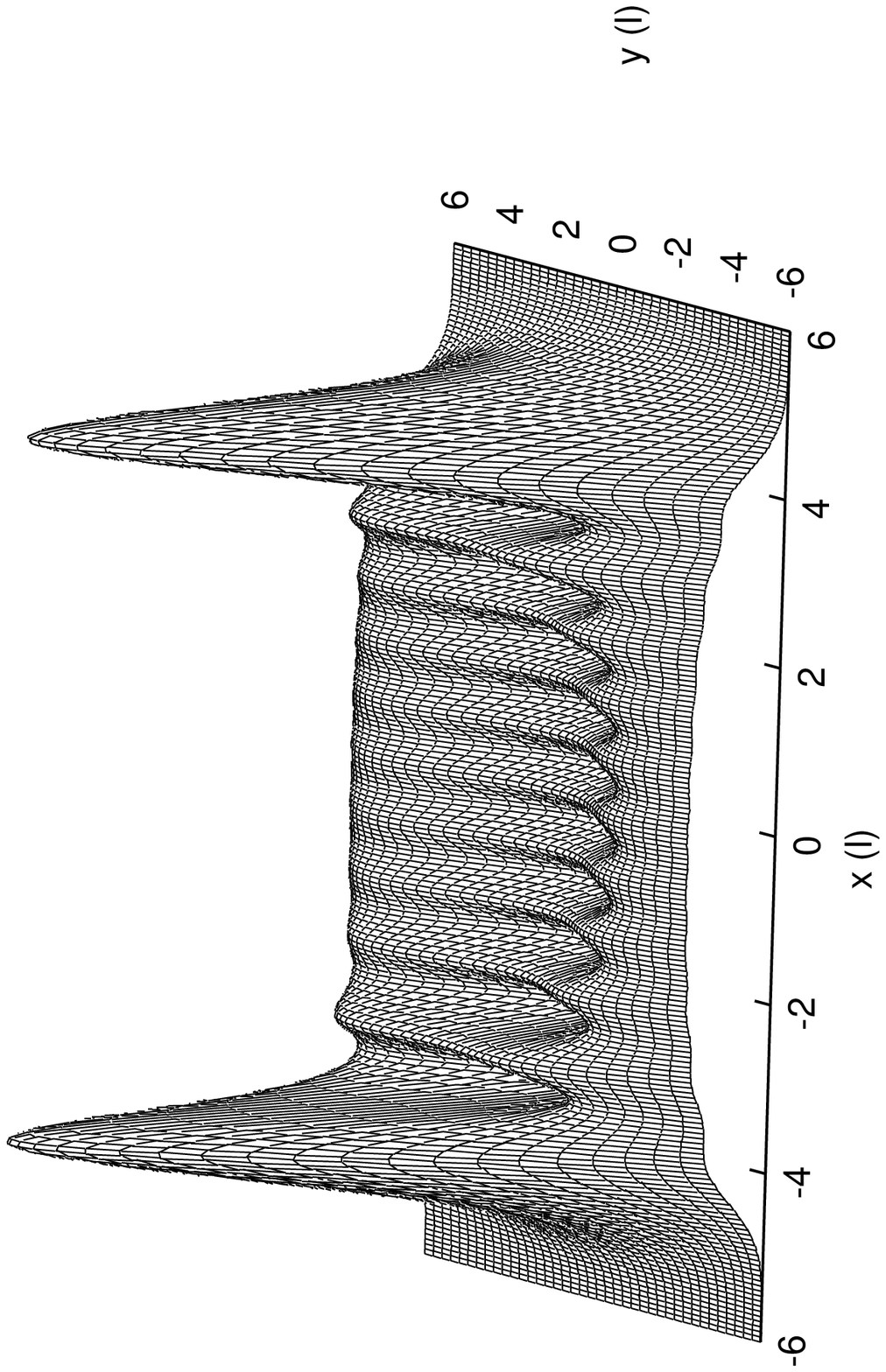}
        \vspace*{0.5cm}
        }
\caption{The probability density for the state $N=10$ ; $|\tanh \Phi_2|
  =0.1, \Phi_{1}= 0$ located at the origin $\bbox{R} = 0$. }
 \label{Fig4}
\end{figure}

Finally, we consider the combined mirror and time-reversal
transformation in Eq.(\ref{Mwia}): ${\cal O}= \sigma_{T}$.  One can
check that $\hat{\sigma }_{T}$ does not change the Hamiltonian in
Eq.(\ref{9aa}) if the symmetric gauge Eq.(\ref{sba}) is chosen.
Ultimately, this is the reason for the mirror symmetry in the
distribution of the density and current seen in Figs(\ref{Fig1}--
\ref{Fig4}).

\newpage


\end{document}